\documentclass{article}
\usepackage{spconf,amsmath,graphicx,hyperref}
\usepackage{amssymb}
\usepackage{booktabs}
\usepackage{multirow}
\usepackage{hyperref}
\usepackage{arydshln}
\usepackage{cite}

\title{MITIGATING ATTENTION SINKS AND MASSIVE ACTIVATIONS IN AUDIO-VISUAL SPEECH RECOGNITION WITH LLMS}
\name{Anand $^{\clubsuit}$ \qquad Umberto Cappellazzo $^{\spadesuit}$ \qquad Stavros Petridis $^{\spadesuit}$ \qquad Maja Pantic $^{\spadesuit}$}
  
  \address{$^{\clubsuit}$ University of British Columbia, Canada \hspace{0.3cm}
      $^{\spadesuit}$ Imperial College London, UK}

\begin{document}
\ninept
\maketitle
\begin{abstract}
Large language models (LLMs) have recently advanced auditory speech recognition (ASR), visual speech recognition (VSR), and audio-visual speech recognition (AVSR). However, understanding of their internal dynamics under fine-tuning remains limited. In natural language processing, recent work has revealed attention sinks, tokens that attract disproportionately high attention, and associated massive activations in which some features of sink tokens exhibit huge activation in LLMs. In this work, we are the first to study these phenomena in multimodal speech recognition. Through a detailed analysis of audio-visual LLMs, we identify attention sinks and massive activations not only at the BOS token but also at intermediate low-semantic tokens across ASR, VSR, and AVSR. We show that massive activations originate in the MLP layers and correspond to fixed feature indices across all sink tokens. We further show that intermediate sink tokens exhibit high cosine similarity to the BOS token, thereby amplifying attention and activation. Building on these insights, we introduce a simple decorrelation loss that reduces cosine similarity between BOS and other tokens, effectively mitigating intermediate sinks and massive activations. Furthermore, our method improves word error rate (WER) under high audio-visual feature downsampling while remaining stable at lower downsampling rates.
\end{abstract}
\section{Introduction}
\label{sec:intro}
Pre-trained Large Language Models (LLMs) have shown remarkable ability to adapt to new domains through parameter-efficient fine-tuning \cite{li2023blip, gao2023llama, fathullah2024prompting, ma2024embarrassingly, maaz2023video, chen2024llast, zang2025contextual}. Recent works demonstrate their effectiveness in Auditory Speech Recognition (ASR), Visual Speech Recognition (VSR), and Audio-Visual Speech Recognition (AVSR) \cite{cappellazzo2025large, thomas2025vallr, yeo2024visual, cappellazzo2025adaptive, yeo2025mms, llama-SMoP, yeo2025zero, cappellazzo2025mome, cappellazzo2025omni}. These approaches extract modality-specific embeddings from pre-trained encoders, downsample them for efficiency, and map them into the LLM embedding space through projectors. The resulting audio and video tokens, concatenated with an instruction prompt, are fed to the LLM, which generates transcriptions autoregressively and is fine-tuned via LoRA \cite{hu2022lora}. Despite these advances, the internal mechanisms underlying audio-visual LLMs remain poorly understood.


Studies of LLMs in NLP and vision have revealed that, within self-attention, the BOS (Beginning of Sentence) special token and certain semantically uninformative intermediate tokens often attract disproportionately large attention \cite{xiao2023streamingllm, gu2024attention, yu2024unveiling, cancedda2024spectral, barbero2025llms, qiu2025gated}. These sink tokens give rise to the phenomenon of attention sinks \cite{xiao2023streamingllm}. While the BOS sink can be useful, acting as key biases that stabilize predictions \cite{gu2024attention} and mitigate forgetting in long contexts \cite{xiao2023streamingllm}, the role of intermediate sinks is less understood. Analyses in NLP suggest they may harm performance \cite{yu2024unveiling}. Moreover, sink tokens exhibit massive activations, where a small subset of hidden-state features reach magnitudes up to four orders larger than the median \cite{sun2024massive, owen2025refined, kaul2024attention}. Yet, the interaction between attention sinks and massive activations remains unclear.


Understanding the internal dynamics of audio-visual LLMs is essential for both interpretability and performance. In this context, BOS sinks may aid performance \cite{xiao2023streamingllm, gu2024attention}, but intermediate sinks risk disrupting the alignment between audio and visual streams by diverting attention from phonetic or lip-movement cues. Similarly, massive activations can over-amplify irrelevant features. Despite their potential impact, these phenomena remain unexplored in speech recognition, leaving a gap in how we understand audio-visual LLMs’ integration of heterogeneous signals.


We present the first extensive analysis of the internal mechanisms of multimodal speech recognition with LLMs. Using Llama-AVSR \cite{cappellazzo2025large}, we reveal the presence of attention sinks at both BOS and intermediate tokens across ASR, VSR, and AVSR tasks. Unlike the BOS sink, which exists in the pre-trained LLM, we observed that intermediate sinks emerge during fine-tuning. We further show that massive activations in these sink tokens originate as early as layer 2 from the MLP component of the transformer block, and that the massively activated feature indices are shared between BOS and intermediate sink tokens. This stems from our key observation that \textit{intermediate sink hidden states exhibit high cosine similarity with the BOS hidden state}.

To probe the role of intermediate sinks, we evaluate performance after mitigating them. We propose a lightweight decorrelation loss that reduces cosine similarity between BOS and other tokens, thereby addressing both attention sinks and massive activations. Prior sink-mitigation strategies, such as prepending placeholder tokens or modifying softmax (e.g., Softmax-off-by-one, SoftPick \cite{xiao2023streamingllm, zuhri2025softpick}), require full pretraining and mainly target BOS sinks in long-context settings. Attention Calibration (ACT) \cite{yu2024unveiling} adjusts attention during inference but adds overhead and does not address massive activations. In contrast, our method integrates seamlessly with LoRA-based fine-tuning, incurs no inference-time cost, and improves WER across ASR, VSR, and AVSR, even under high compression.

Our key contributions are: \textbf{(1)} We provide the first analysis of attention sinks and massive activations in audio-visual LLMs across ASR, VSR, and AVSR. \textbf{(2)} We identify the origin of massive activations and explain the co-existence of massive activations and attention sinks via cosine similarity. \textbf{(3)} We introduce a novel decorrelation loss that mitigates intermediate sinks and massive activations while improving WER at high compression rates.

\section{Preliminaries}
\label{sec:preliminaries}
\textbf{Llama-AVSR.} We begin our analysis by revisiting the architecture of Llama-AVSR \cite{cappellazzo2025large}, which forms the foundation of our study. In this setting, raw audio and video inputs are first encoded into modality-specific embeddings using pre-trained encoders. In our setting, we use AV-HuBERT \cite{shi2022learning} as video encoder and Whisper \cite{radford2023robust} as the audio encoder. Since these embeddings are high-dimensional and temporally dense, directly feeding them into the LLM would be computationally expensive. To address this, Llama-AVSR applies a compression step that temporally downsamples the embeddings via average pooling, before projecting them into the LLM embedding space using lightweight linear projectors. We denote compression rates as $(a,v)$ for AVSR, where $a$ and $v$ are the downsampling factors for audio and video tokens respectively (e.g., AVSR (16,5)), and as a single value $(a)$ or $(v)$ for unimodal ASR and VSR. The compressed audio $\mathbf{X}_{\text{aud}}$ and video $\mathbf{X}_{\text{vid}}$ tokens are then concatenated with an instruction prompt $\mathbf{X}_{\text{inst}}$ and passed to the LLM, which autoregressively generates the target output $\mathbf{Y}$ as:
\begin{equation}
p(\mathbf{Y}|\mathbf{X}_{\text{aud}}, \mathbf{X}_{\text{vid}}, \mathbf{X}_{\text{inst}})
= \prod_{i=1}^{N} p(y_i|\mathbf{X}_{\text{aud}}, \mathbf{X}_{\text{vid}}, \mathbf{X}_{\text{inst}}, y_{<i}),
\end{equation}
where $N$ is number of tokens and $y_{<i}$ is the generated output sequence up to token $i-1$.
\newline
\newline
\noindent\textbf{Autoregressive LLMs.} Autoregressive LLMs are typically constructed by stacking $L$ transformer decoder blocks \cite{vaswani2017attention}. Each block consists of a Multi-Head Self Attention (MHSA) module followed by a Multilayer Perceptron (MLP). Given hidden states $\textbf{H}^{l-1}\in\mathbb{R}^{N\times d}$ at layer $l-1$, MHSA computes pair-wise relationships between the tokens with help of queries $\textbf{Q}_h^l$, keys $\textbf{K}_h^l$, and values $\textbf{V}_h^l$ computed for each head $h$ from linear projection of each token's $d$-dimensional hidden state. The attention map is then computed with:
\begin{equation}
    \textbf{A}_h^l = \text{Softmax}\left(\frac{\textbf{Q}_h^l{\textbf{K}_h^l}^\top}{\sqrt{d_h}} + \textbf{M}\right),
\end{equation}
where $d_h=d/H$ where $H$ is number of heads and $\textbf{M}\in\mathbb R^{N\times N}$ is the causal mask. The head outputs $\textbf{O}^l_h = \textbf{A}^l_h \textbf{V}^l_h$, are concatenated and projected to get the output $\mathbf{O}^l$ of MHSA. In pre-norm LLM blocks, the hidden state for next layer is then computed by performing MHSA on Layer Normalized (LN) hidden state $\textbf{H}^{l-1}$ as:
\begin{equation}
    \textbf{H}^{l} = \textbf{H}^{l-1} + \mathbf{O}^l + \text{MLP}(\text{LN}(\textbf{H}^{l-1} + \mathbf{O}^l)).
\end{equation}
\begin{figure}[htb]
\begin{minipage}[b]{.48\linewidth}
  \centering
  \centerline{\includegraphics[width=4.0cm]{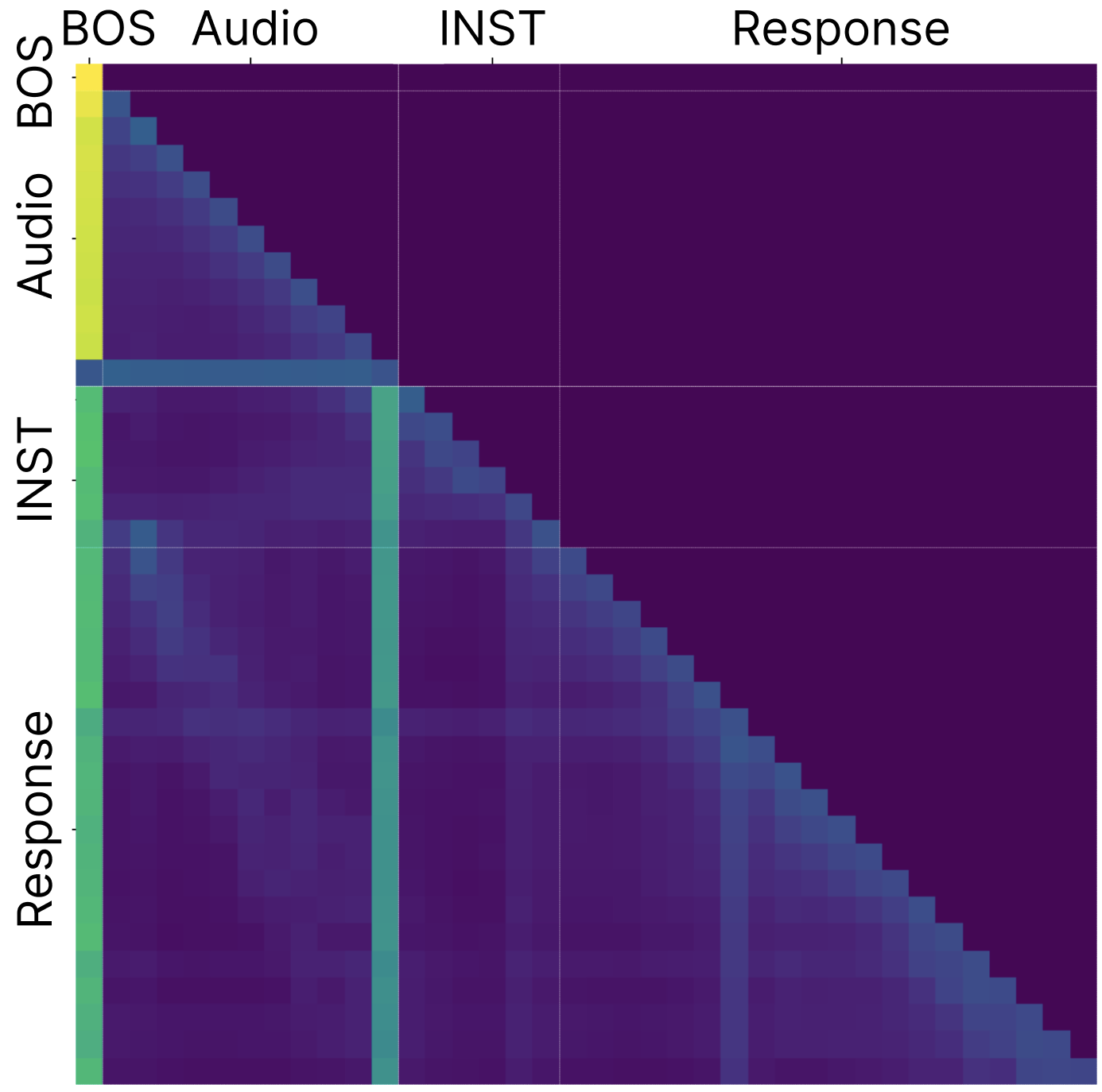}}
  \centerline{(a) Llama-ASR (32)}\medskip
\end{minipage}
\hfill
\begin{minipage}[b]{0.48\linewidth}
  \centering
  \centerline{\includegraphics[width=4.38cm]{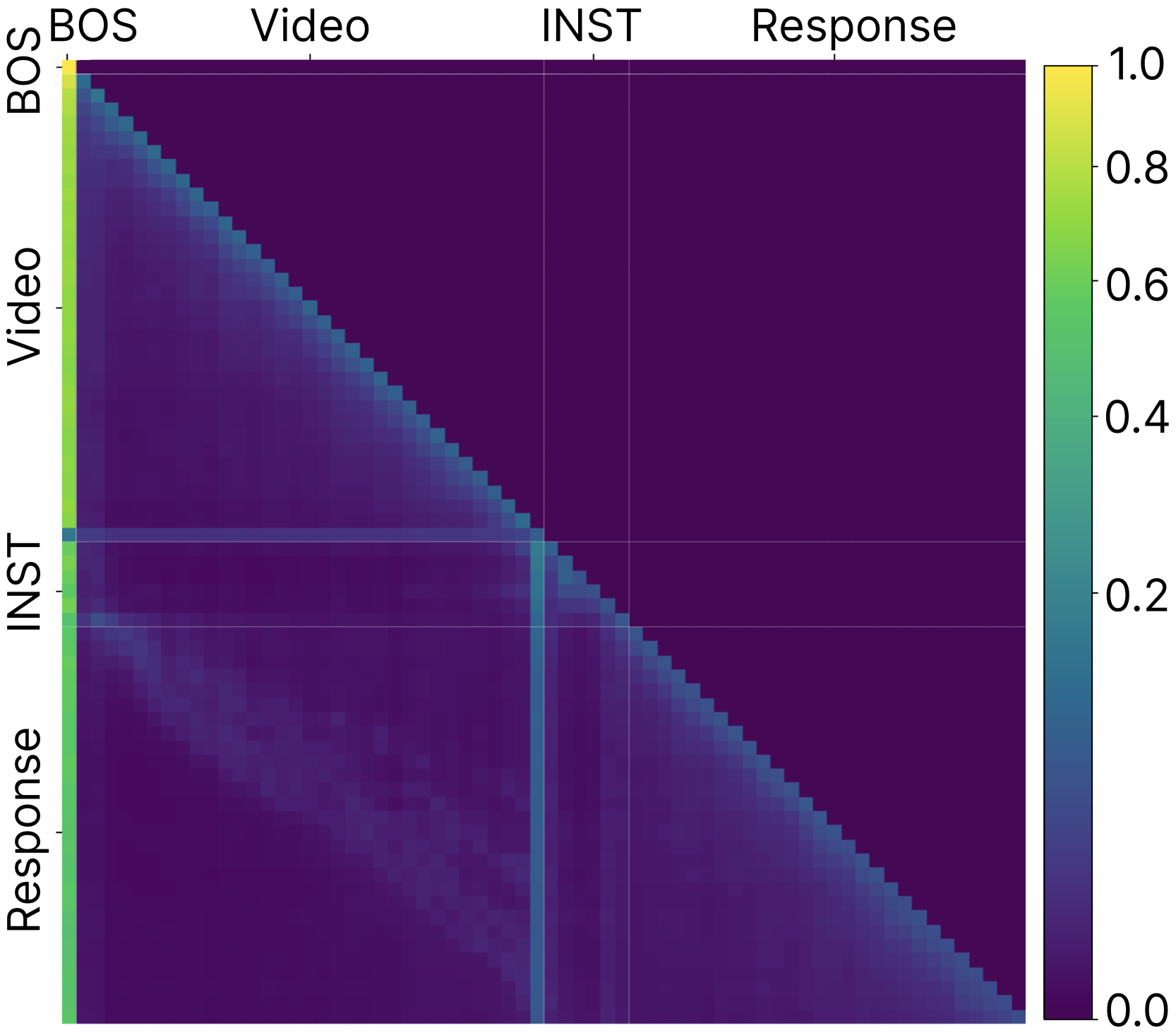}}
  \centerline{(b) Llama-VSR (5)}\medskip
\end{minipage}
\begin{minipage}[b]{.48\linewidth}
  \centering
  \centerline{\includegraphics[width=4.0cm]{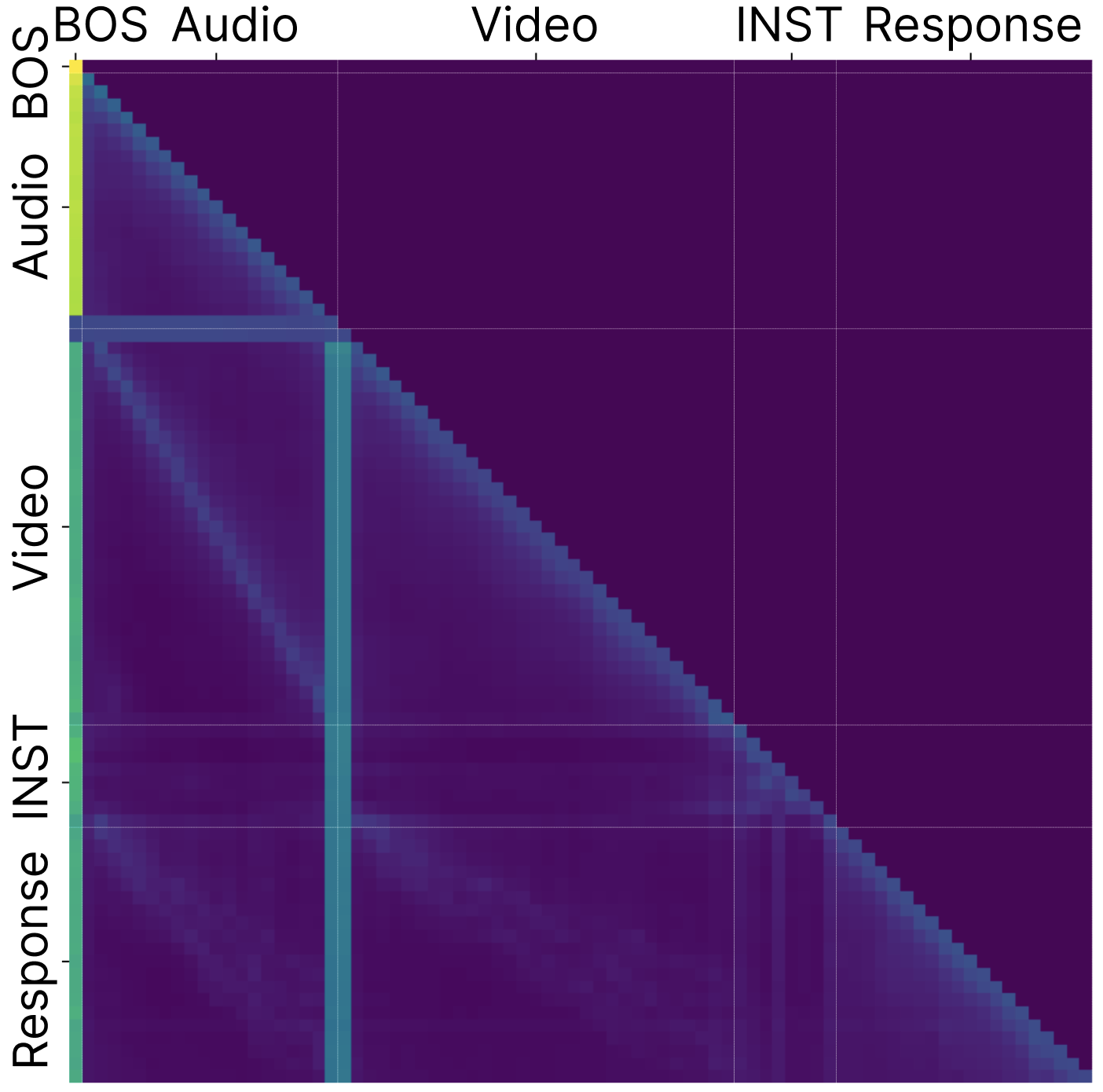}}
  \centerline{(c) Llama-AVSR (16, 5)}\medskip
\end{minipage}
\hfill
\begin{minipage}[b]{0.48\linewidth}
  \centering
  \centerline{\includegraphics[width=4.38cm]{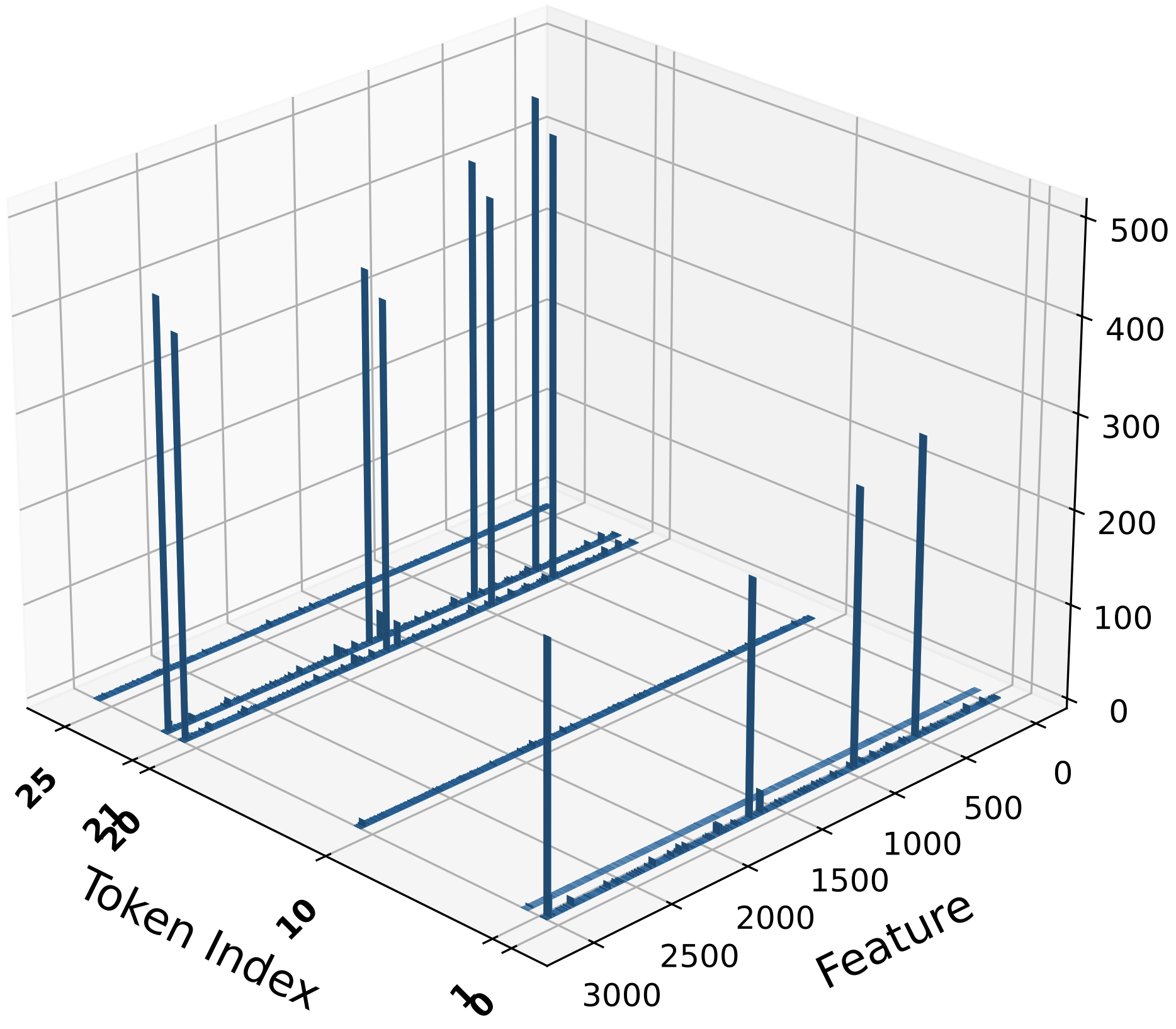}}
  \centerline{(d) Llama-AVSR (16,5)}\medskip
\end{minipage}
\caption{\textbf{(a,b,c)} Attention sinks present in BOS and intermediate tokens for different tasks and compression rates (e.g., AVSR task at audio-video compression rates of (16, 5)). \textbf{(d)} Activation magnitudes (z-axis) of the hidden state in Llama-AVSR (16, 5) at layer 5 reveal some features with massive activation in sink tokens.}
\label{fig:attention_map}
\vspace{-0.3cm}
\end{figure}
The MLP processes each token independently, often in a gated linear unit (GLU) form as follows:
\begin{equation}
    \text{MLP}(h) = ((h\textbf{W}_{\text{up}}) \odot 
    \sigma(h\textbf{W}_{\text{gate}}))\textbf{W}_{\text{down}},
\end{equation}
where $\sigma$ is a non-linear activation function and $\textbf{W}_{\text{gate}}$, $\textbf{W}_{\text{up}}$, $\textbf{W}_{\text{down}}\in\mathbb R^{d\times d'}$ and $\odot$ is element-wise product.

\section{Analysis of Audio-Visual LLMs}
\label{sec:analysis}
\subsection{Attention Sinks}
\label{subsec:sink}
\begin{figure}[htb]
\begin{minipage}[b]{.48\linewidth}
  \centering
  \centerline{\includegraphics[width=4.0cm]{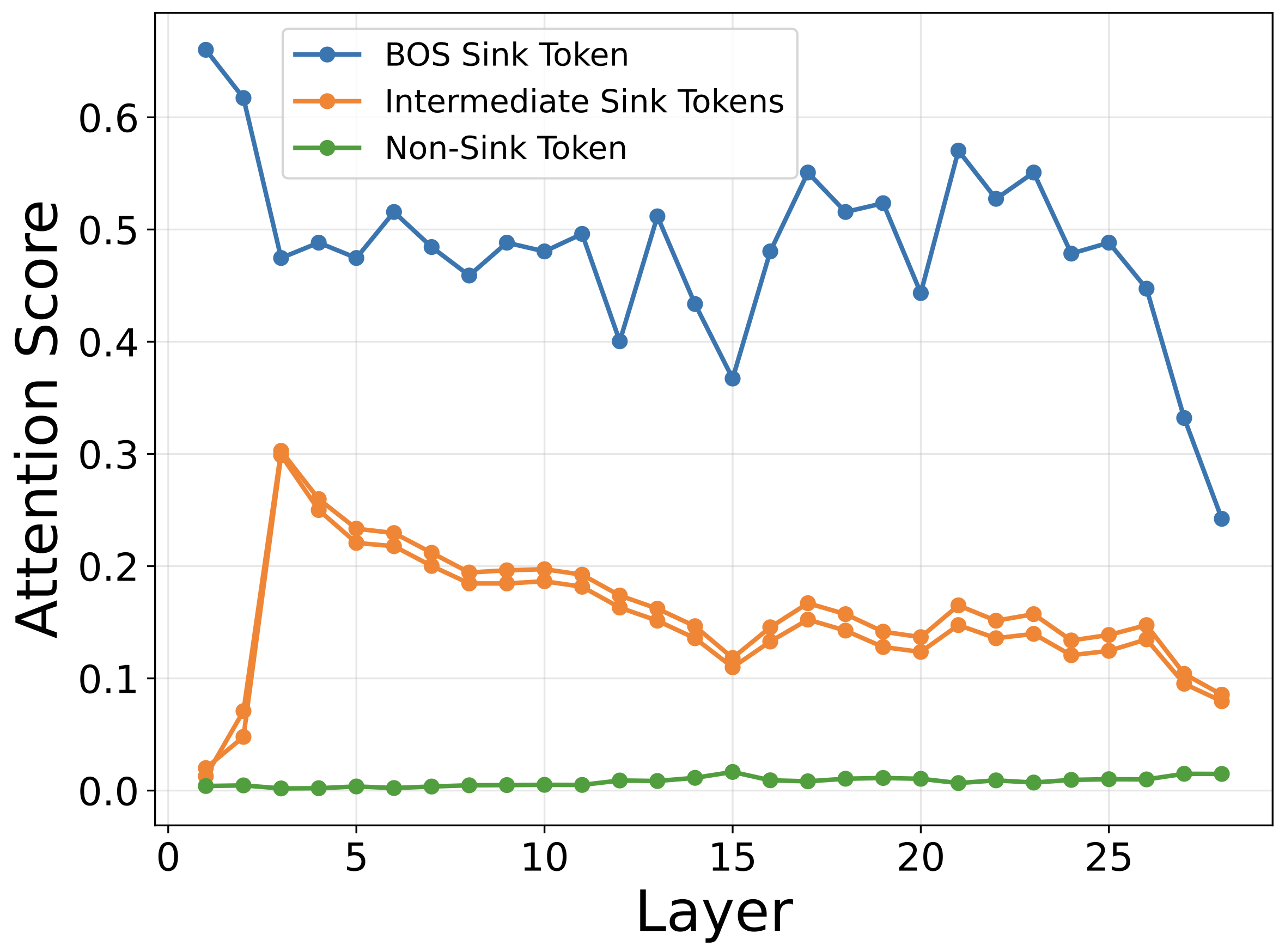}}
  \centerline{(a)}\medskip
\end{minipage}
\begin{minipage}[b]{.48\linewidth}
  \centering
  \centerline{\includegraphics[width=4.0cm]{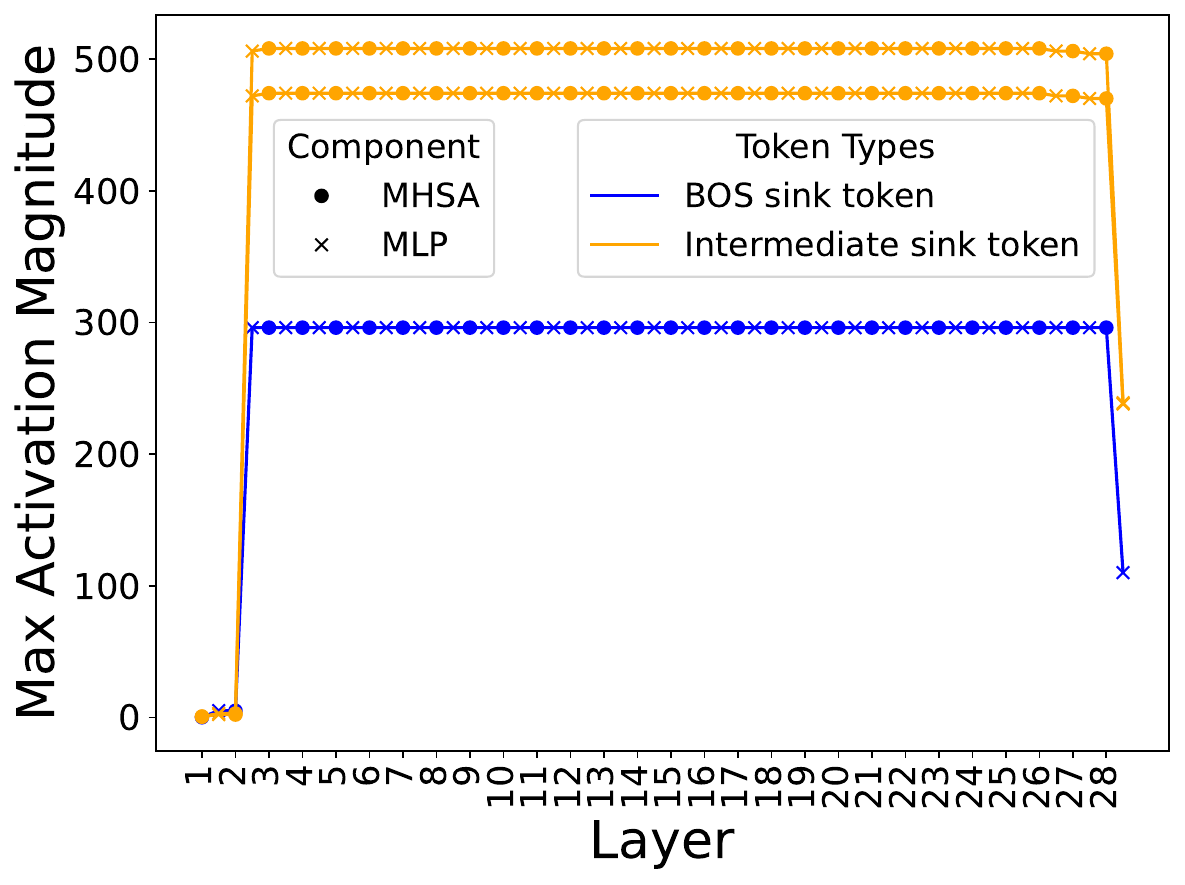}}
  \centerline{(b)}\medskip
\end{minipage}
\caption{\textbf{(a)} Intermediate attention sinks originate after layer 2 in Llama-AVSR (16,5). \textbf{(b)} Massive activations in Llama-AVSR (16,5) originate from the MLP of Layer 2.}
\label{fig:layer_analysis}
\vspace{-0.5cm}

\end{figure}

Each element of the attention map $\textbf{A}_h^l\in\mathbb R^{N\times N}$ given by $\textbf{A}_h^l[i,j]$ represents the attention token $i$ gives to token $j$. The causal mask $\textbf{M}$ ensures $\textbf{A}_h^l[i,j] = 0$ for all $i < j$. Thus, we compute attention score of token $i$ at layer $l$ as average attention it receives from other tokens across all the heads as
\begin{equation}
\alpha_i^l := \frac{1}{H(N-i+1)}\sum_{h=1}^H\sum_{k=1}^N \textbf{A}_h^l[k,i].
\end{equation}
We perform our analysis on Llama-based LLMs \cite{dubey2024llama} (since the trend is similar across LLMs, we report the results with Llama 3.2-3B) on ASR, VSR, and AVSR tasks with different compression rates and compute the attention scores $\alpha_i^l$ for each token across all layers. Figure~\ref{fig:attention_map} (a–c) presents the average attention maps aggregated across all heads and layers. We observe that the initial token consistently receives substantially higher attention compared to other tokens across layers, confirming the presence of a BOS sink. This observation is consistent with prior work \cite{xiao2023streamingllm}, which demonstrated that the BOS token serves as an attention sink. Furthermore, after layer 2, certain intermediate tokens begin to attract elevated attention scores as shown in Figure~\ref{fig:layer_analysis}(a), suggesting the emergence of intermediate attention sinks. These patterns highlight both BOS and intermediate sinks in ASR, VSR, and AVSR under different compression rates. Computing the attention score at different epochs during fine-tuning showed that these intermediate sinks appear as a result of fine-tuning unlike the BOS sink which we noticed is already present in the pre-trained LLM. 

Finally, we also notice that the intermediate sink tokens occur on tokens with low semantic value. This includes special tokens like \texttt{<audio>}, \texttt{</audio>}, \texttt{<video>}, \texttt{</video>} and prompt tokens. We hypothesize that this phenomenon occurs because these tokens are consistently present during training, leading the LLM to use them as anchors to absorb excessive attention while optimizing the loss function. Under high compression, where informative audio–visual tokens are scarce, intermediate sinks may bias attention toward semantically weak tokens at the expense of phonetic and lip-movement cues, potentially contributing to degraded performance. Later, in Section~\ref{sec:results}, we observe that mitigating these intermediate sinks leads to improved WER at high compression rates.

\subsection{Massive Activations}
\label{subsec:activation}
In addition to attention sinks, we analyze \textit{massive activations}, a phenomenon where certain hidden-state features of a token exhibit extremely large magnitudes compared to the median \cite{sun2024massive}, despite the presence of normalization layers. Formally, for token $i$ at layer $l$, we define the massively activated feature index set as:
\begin{equation}
\Theta_i^l := \left\{ j \in \{1, \dots, d\} \;\middle|\; |\textbf{H}^l[i,j]| \ge \tau \cdot \text{median}(|\textbf{H}^l|) \right\},
\end{equation}
where $\textbf{H}^l[i,j]$ denotes the $j$-th feature of the hidden state of token $i$ at layer $l$, and $\text{median}(|\textbf{H}^l|)$ is computed over the magnitudes of all features across all tokens in layer $l$. For our analysis, we set $\tau = 10^3$.

Analyzing $\Theta_i^l$ across multiple LLMs trained for ASR, VSR, and AVSR on different compression rates, we empirically observed that $\Theta_i^l$ is non-empty if and only if $i$ is a sink token and $l\in \{2,3,\ldots,L-1\}$ i.e., massive activations do not happen in first and last layer of the LLM.  \textit{This suggests that the phenomena of attention sinks and massive activations co-exist in intermediate layers of the LLM}. Moreover, we found that $\Theta_i^l$ is identical for all the sink tokens. Figure~\ref{fig:attention_map}(d) shows presence of massive activations in all 3 sink tokens with $\Theta_0^l = \Theta_{20}^l = \Theta_{21}^l$ in layer 5 of LLM as sinks were present at token indices $\{0,20,21\}$.

To further investigate the origin of massive activations, we analyzed the contributions of different components in layer 2 and found that it arises from the MLP module as shown in Figure~\ref{fig:layer_analysis}(b). Specifically, in layer 2 of LLaMA3.2-3B, we observed that within the GLU, the term $h\textbf{W}_{\text{gate}}$ exhibits large positive values for a fixed set of features across all sink tokens, and negative values for non-sink tokens. Due to the non-linear activation function $\sigma$ (typically SiLU), only the positive values attain high magnitudes. These are further amplified by the element-wise product with $h\textbf{W}_{\text{up}}$, resulting in massive activations in the MLP latent space $\mathbb{R}^{d'}$. This amplified signal is then propagated to the LLM’s hidden state, producing the observed massive activations in sink tokens through down projection of MLP.

\subsection{Cosine Similarity with BOS}
\label{subsec:cosine_sim}
\begin{figure}[t]
\begin{minipage}[b]{.48\linewidth}
  \centering
  \centerline{\includegraphics[width=4.38cm]{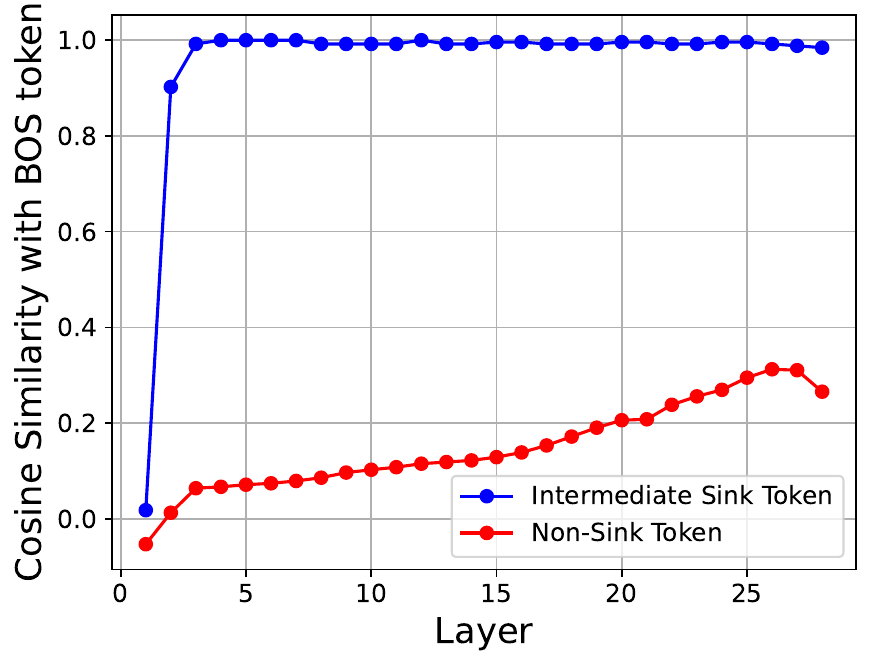}}
  \centerline{(a)}\medskip
\end{minipage}
\hfill
\begin{minipage}[b]{0.48\linewidth}
  \centering
  \centerline{\includegraphics[width=4.38cm]{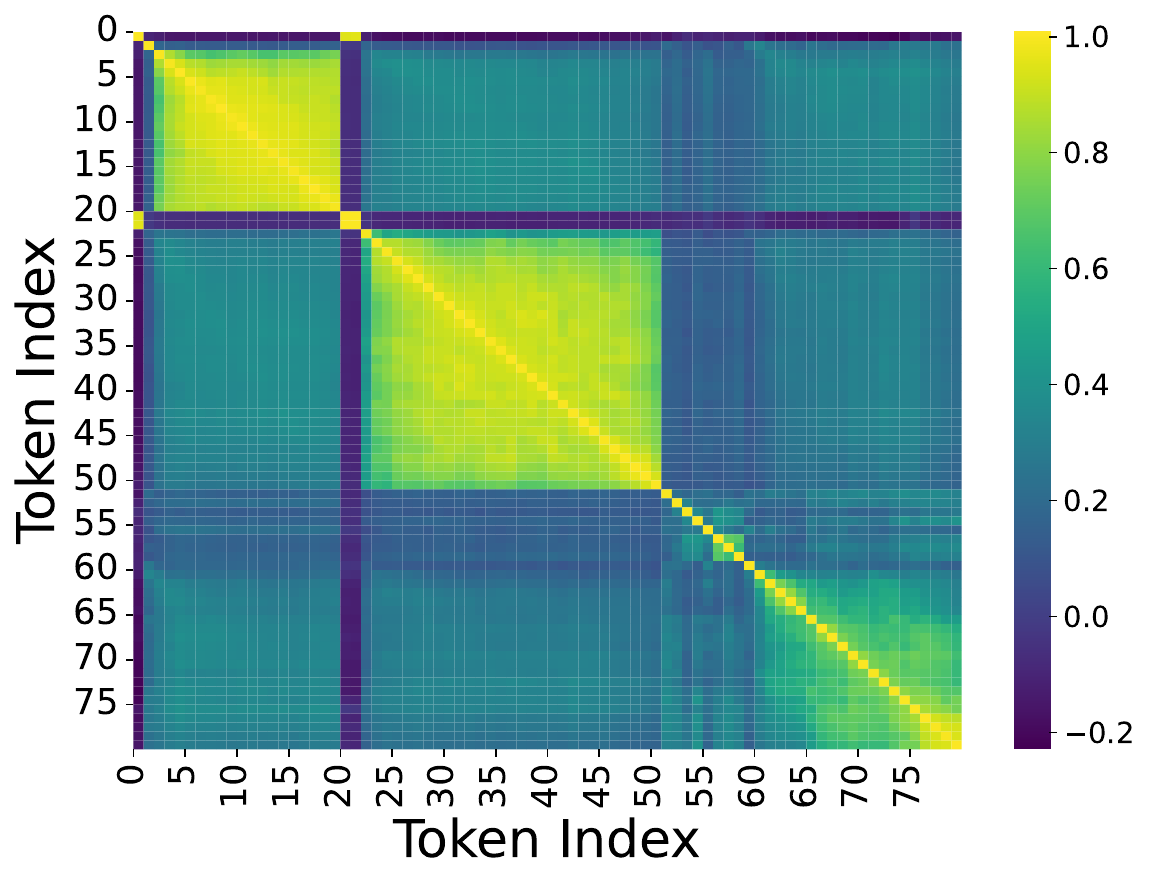}}
  \centerline{(b)}\medskip
\end{minipage}
\caption{\textbf{(a)} Cosine similarity of intermediate sink tokens and non-sink tokens with BOS token across layers of Llama-AVSR (16, 5). \textbf{(b)} Pairwise cosine similarity heatmap of hidden-state embeddings of tokens of Llama-AVSR (16, 5) in layer 5.}
\label{fig:cosine_sim}
\vspace{-0.5cm}
\end{figure}

To understand why intermediate sink tokens share the same set of massively activated features as the BOS token and the co-existence of massive activations and attention sinks, we analyzed the directional similarity of their hidden states. Specifically, we computed the cosine similarity between the hidden state of each intermediate sink token $\textbf{H}^l[i,:]$ and that of the BOS token $\textbf{H}^l[0,:]$ across layers:
\begin{equation}
\text{cos-sim}(\textbf{H}^l[i], \textbf{H}^l[0]) = \frac{\textbf{H}^l[i] \cdot \textbf{H}^l[0]}{\|\textbf{H}^l[i]\|_2 \|\textbf{H}^l[0]\|_2}.
\end{equation}
We observe that hidden states of intermediate sink tokens are highly aligned with the BOS token from layer 2 onwards as shown in Figure~\ref{fig:cosine_sim}(a). This directional alignment explains why intermediate sink tokens exhibit identical massively activated feature indices $\Theta_i^l$ and receive excessive attention; as their hidden states point in nearly the same direction as the BOS token, they activate the same set of features and inherit the same attention patterns. \textit{These findings indicate that the root cause of both attention sinks and massive activations in intermediate tokens is the alignment of their hidden states with the BOS token with high cosine similarity}. Figure~\ref{fig:cosine_sim}(b) illustrates the cosine similarity between tokens in Llama-AVSR (16, 5). Sink tokens, located at indices $\{0,20,21\}$, exhibit very high pairwise cosine similarity, indicating that their hidden states are closely aligned. In contrast, all other tokens show orthogonal behavior with these sink tokens, suggesting that the directional alignment is specific to sink tokens and does not extend to regular tokens.

To test whether BOS alignment drives attention sinks and massive activations, we perform controlled rotations of tokens. Specifically, for an intermediate sink token $i$, we rotate its hidden state towards the nearest non-sink token $f(i)$ as:
\begin{equation}
    \textbf{H}^l[i] \leftarrow \|\textbf{H}^l[i]\|_2 \cdot \frac{\textbf{H}^l[f(i)]}{\|\textbf{H}^l[f(i)]\|_2}.
\end{equation}
\begin{figure}[htb]
\begin{minipage}[b]{.48\linewidth}
  \centering
  \centerline{\includegraphics[width=4.0cm]{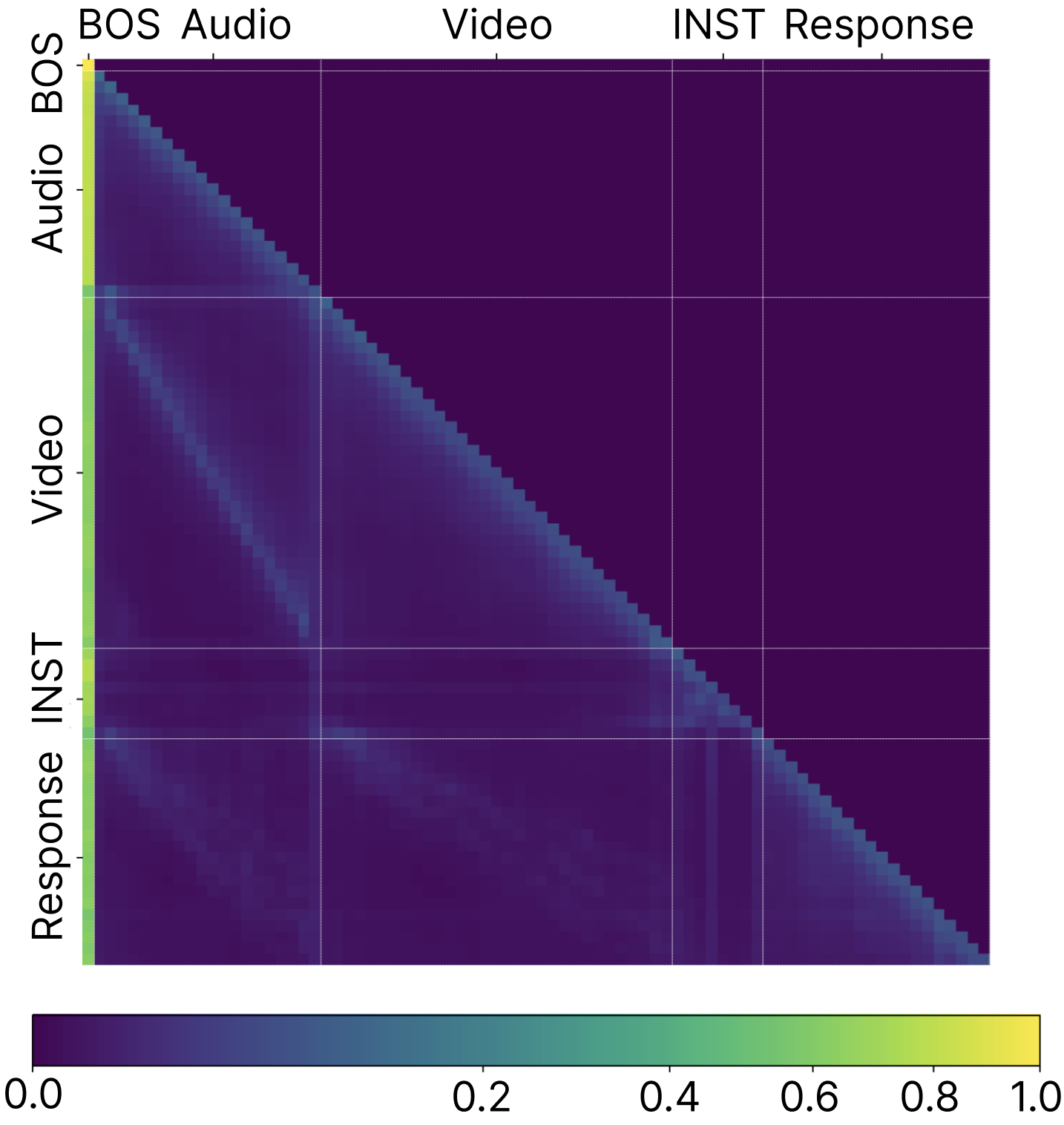}}
  \centerline{(a)}\medskip
\end{minipage}
\hfill
\begin{minipage}[b]{0.48\linewidth}
  \centering
  \centerline{\includegraphics[width=4.38cm]{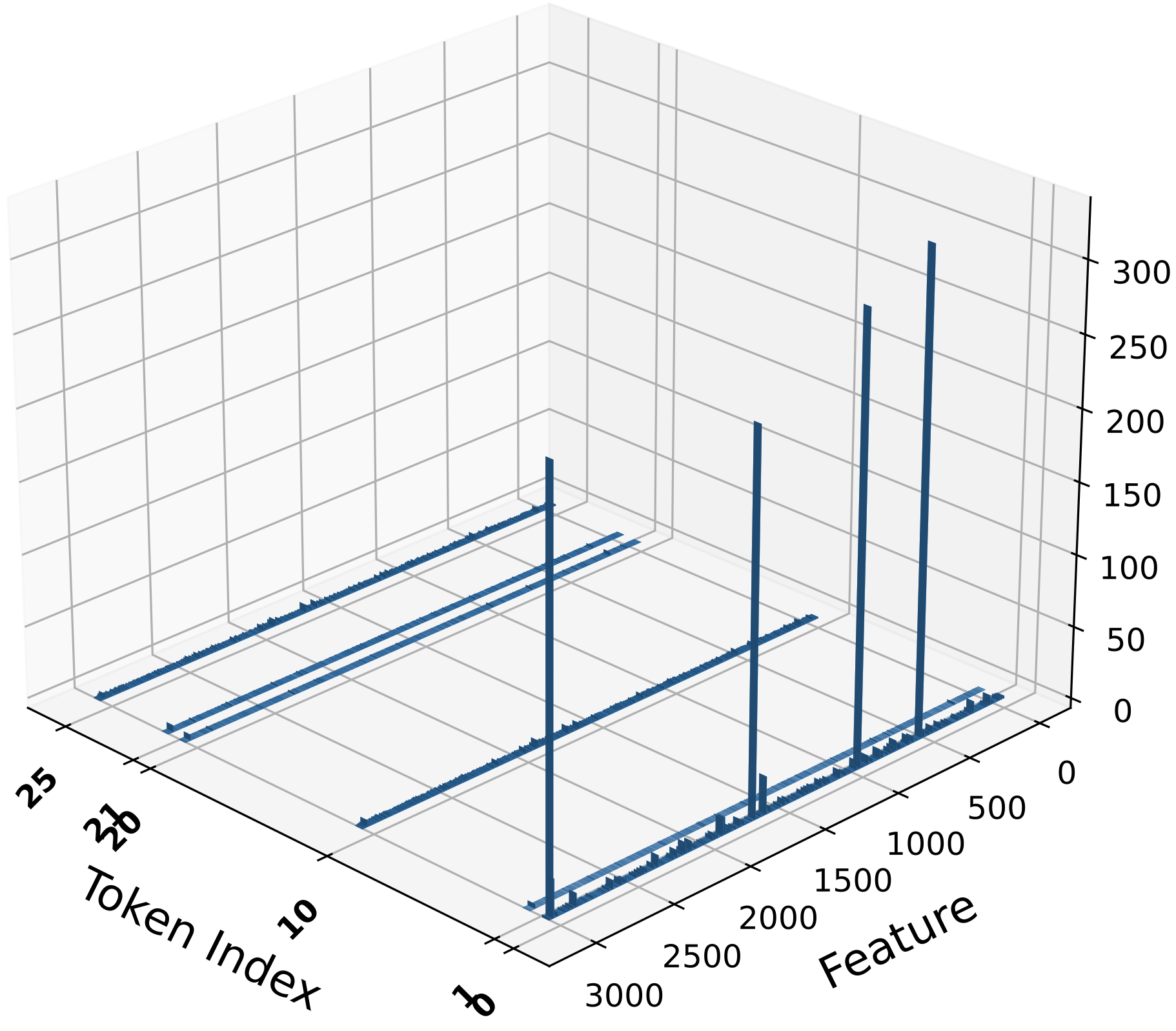}}
  \centerline{(b)}\medskip
\end{minipage}
\begin{minipage}[b]{.48\linewidth}
  \centering
  \centerline{\includegraphics[width=4.0cm]{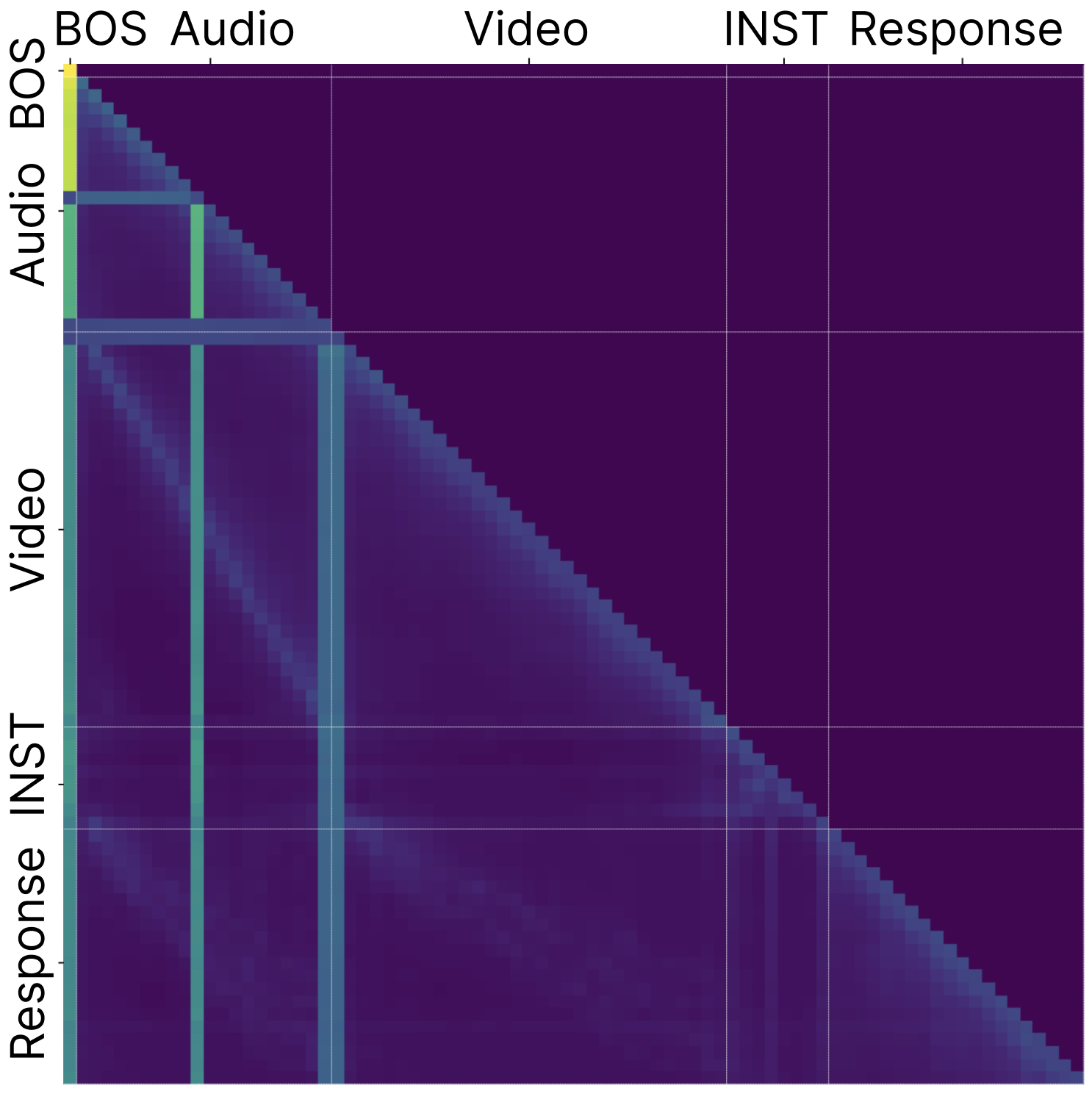}}
  \centerline{(c)}\medskip
\end{minipage}
\hfill
\begin{minipage}[b]{0.48\linewidth}
  \centering
  \centerline{\includegraphics[width=4.38cm]{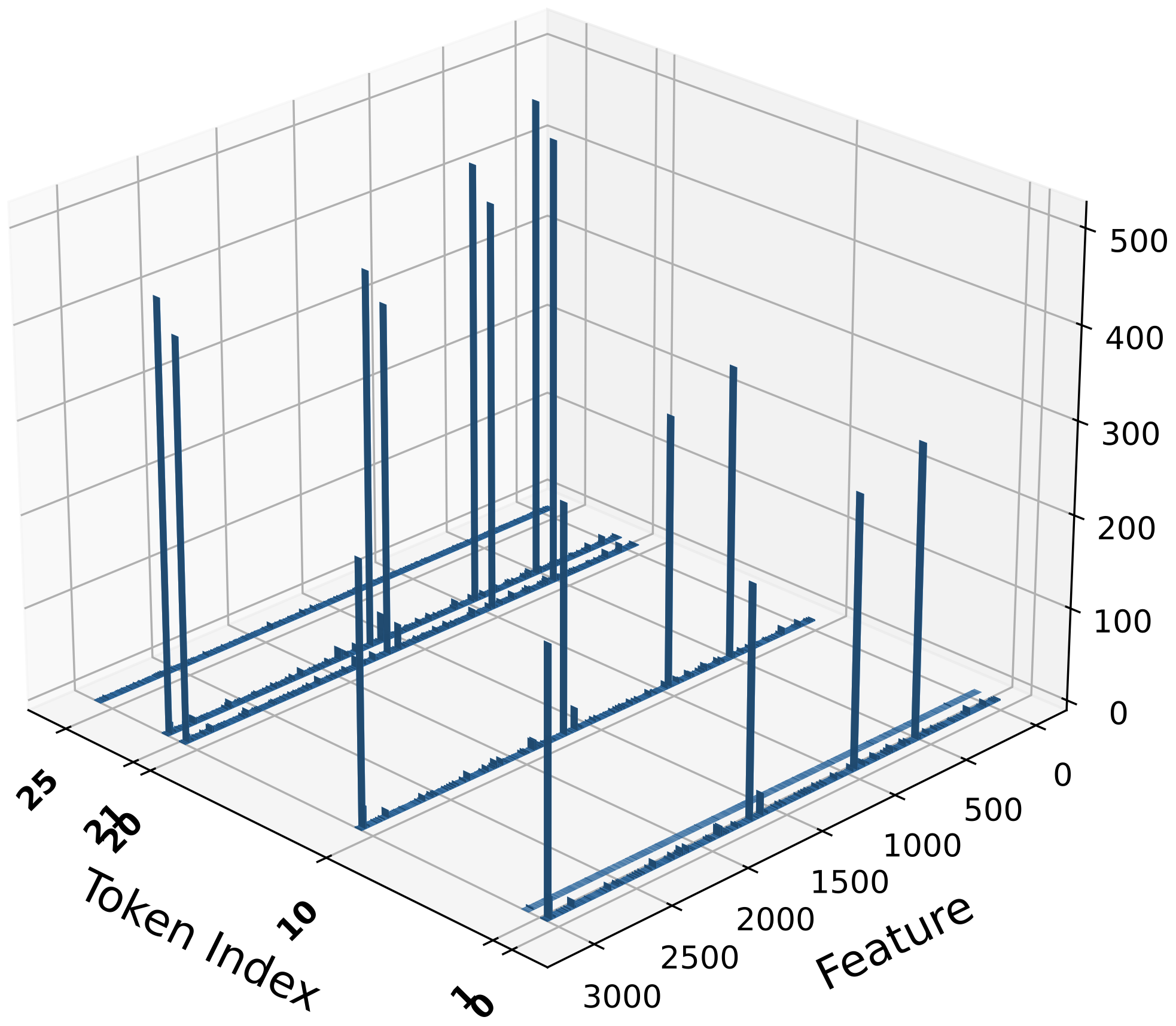}}
  \centerline{(d)}\medskip
\end{minipage}
\caption{\textbf{(a,b)} Sinks and massive activations vanish when disaligned from BOS. \textbf{(c,d)} They emerge when aligned with BOS.}
\label{fig:sink_rotate}
\vspace{-0.3cm}
\end{figure}
Applied to sink tokens at indices $\{20,21\}$ in Llama-AVSR (16, 5), this operation removes both attention sinks and massive activations, as shown in Figure~\ref{fig:sink_rotate}(a,b). Conversely, when we rotate a non-sink token towards the BOS token direction,
\begin{equation}
    \textbf{H}^l[i] \leftarrow \|\textbf{H}^l[i]\|_2 \cdot \frac{\textbf{H}^l[0]}{\|\textbf{H}^l[0]\|_2},
\end{equation}
we observe the emergence of attention sink behavior and massive activations at that position as shown in Figure~\ref{fig:sink_rotate}(c,d) where a sink emerged at index 10.

\section{Proposed Method}
\label{sec:method}
Our analysis in Section~\ref{subsec:cosine_sim} reveals that the root cause of both attention sinks and massive activations in intermediate tokens is their directional alignment with the BOS token in hidden-state space. While these phenomena emerge naturally during training, whether their existence is beneficial for model's performance is a natural question. Motivated by this, we propose a simple yet effective \textbf{decorrelation loss} that explicitly discourages alignment between the BOS token and other tokens, thereby mitigating both attention sinks and massive activations in intermediate tokens.

\subsection{Decorrelation Loss}
Let $\mathbf{H}^l \in \mathbb{R}^{N \times d}$ denote the hidden states at layer $l$. Our decorrelation loss penalizes similarity between BOS and non-BOS tokens:
\begin{equation}
    \mathcal{L}_{\text{decor}} = \frac{1}{(N-1)(L-2)} \sum_{l=2}^{L-1}\sum_{i=1}^{N-1} 
    \text{cos-sim}(\textbf{H}^l[i], \textbf{H}^l[0])^2.
\end{equation}
We exclude the first and last layers, where massive activations do not occur. Squaring the cosine similarity yields smoother gradients and stronger penalties for highly aligned tokens.

\subsection{Final Training Loss}
We combine the decorrelation loss with the standard cross-entropy loss for autoregressive speech recognition:
\begin{equation}\label{eqn:method}
    \mathcal{L} = \mathcal{L}_{\text{CE}} + \lambda \cdot \mathcal{L}_{\text{decorr}},
\end{equation}
where $\lambda$ is a hyperparameter that controls the strength of decorrelation regularization. We report the results of our experiments by selecting the value of $\lambda \in \{10, 10^2, 10^4\}$ that yields the best performance. Importantly, our method requires no modification to the model architecture and adds negligible computational overhead, as cosine similarity is computed directly from hidden states.

\section{Experimental Results}
\label{sec:results}
Based on Equation~\ref{eqn:method}, we investigate whether the removal of the intermediate attention sinks and massive activations help the model in terms of WER performance. All our experiments follow the training details and the code provided in \cite{cappellazzo2025large}. We report the results using Llama 3.2-3B \cite{dubey2024llama} as LLM. We observed similar trends for other LLMs (i.e., Llama 3.2-1B and Llama 2-7B).
\subsection{Decorrelation Loss for Intermediate Attention Sinks}
By penalizing BOS alignment, the decorrelation loss encourages intermediate tokens to occupy distinct representational directions in hidden-state space than the inital token. Using our proposed decorrelation loss, we successfully mitigated both attention sinks and massive activations from intermediate tokens. We conducted extensive experiments on the LRS2 dataset for both AVSR and ASR tasks. For VSR, we opted for the LRS3 dataset due to the increased challenge of the task. As shown in Table~\ref{tab:decor}, we observe consistent improvements in WER at high compression rates, while performance remains comparable to the baseline at lower compression rates.
\begin{table}[t]
    \centering
    \caption{Llama-AVSR WER (\%) results with and without decorrelation loss.}
    \begin{tabular}{lcccr}
        \toprule
        \textbf{Task} & \textbf{Compression} & \textbf{Base} & \textbf{Decorr.} & $\Delta$ \\
        \midrule
        \multirow{3}{*}{ASR} 
            & (4)     & 2.62 & 2.61 & +0.01 \\
            & (16)    & 4.83 & 3.91 & +0.92 \\
            & (32)    & 12.92 & 11.50 & +\textbf{1.42} \\
        \hdashline
        \multirow{2}{*}{VSR} 
            & (1)     & 25.84 & 25.63 & +0.21 \\
            & (5)     & 45.19 & 34.08 & +\textbf{11.11} \\
        \hdashline
        \multirow{4}{*}{AVSR} 
            & (1,1)   & 2.26 & 2.22 & +0.04 \\
            & (4,2)   & 2.44 & 2.42 & +0.02 \\
            & (16,5)  & 4.15 & 3.72 & +\textbf{0.43} \\
        \bottomrule
    \end{tabular}
    \label{tab:decor}
    \vspace{-0.2cm}
\end{table}
These results demonstrate that intermediate attention sinks are detrimental to model robustness under compression, and that decorrelation loss provides an effective and lightweight solution.

\subsection{Comparison with Prior Sink Mitigation Methods}
Most prior sink mitigation methods target only the BOS sink token in streaming applications with very long context windows and require full model pre-training \cite{xiao2023streamingllm, zuhri2025softpick}, making them incompatible with our LoRA-based setting. Another approach, Attention Calibration (ACT) \cite{yu2024unveiling}, focuses on intermediate sink mitigation in NLP tasks by redistributing attention in some selected attention heads. We applied ACT on Llama-AVSR (16,5) and ASR (32) settings on LRS2. Unlike in NLP tasks, we observe only marginal improvements for audio-visual speech recognition tasks. Additionally, ACT fails to mitigate massive activations in sink tokens.
\begin{table}[t]
    \centering
    \caption{Comparison between our proposed method and ACT \cite{yu2024unveiling}.}
    \begin{tabular}{lcccc}
        \toprule
        \textbf{Task} & \textbf{Compression} & \textbf{Base} & \textbf{ACT} & \textbf{Decorr. (Ours)} \\
        \midrule
        ASR & (32) & 12.92 & 12.81 & \textbf{11.50} \\
        \hdashline
        AVSR & (16,5) & 4.15 & 4.08 & \textbf{3.72} \\
        \bottomrule
    \end{tabular}
    \label{tab:act_comparison}
    \vspace{-0.3cm}
    
\end{table}
Table~\ref{tab:act_comparison} summarizes the comparison, highlighting that our method effectively addresses both attention sinks and massive activations while providing larger performance gains under high compression rates.


\section{Conclusion}
In this work, we presented the first study of attention sinks and massive activations in multimodal speech recognition LLMs. Our analysis revealed that these phenomena arise from the directional alignment of intermediate tokens with the BOS token. To address this, we proposed a lightweight decorrelation loss that mitigates both effects without architectural changes. Experiments across AVSR, ASR, and VSR showed consistent WER improvements under high compression, demonstrating the effectiveness of our approach.

\clearpage

\bibliographystyle{IEEEbib}
\bibliography{strings,refs}

\end{document}